\documentclass[
preprint,
showpacs,
showkeys
]{revtex4-1}
\usepackage{amsthm}
\theoremstyle{plain}
\newtheorem{theorem}{Theorem}
\newtheorem*{theorem*}{Theorem}

\newtheorem*{definition*}{Definition}

\newtheorem*{lemma*}{Lemma}

\newtheorem{remark}[theorem]{Remark}

\usepackage{braket}
\usepackage{delarray}
\usepackage{here}

\usepackage{txfonts}

\usepackage[dvipdfmx]{graphicx}
\usepackage{bm}
\usepackage{color}
\usepackage{ulem}

\newcommand{\be}{\begin{eqnarray}}
\newcommand{\ee}{\end{eqnarray}}
\newcommand{\ba}{\begin{array}}
\newcommand{\ea}{\end{array}}
\newcommand{\bmat}{\left(\begin{array}}
\newcommand{\emat}{\end{array}\right)}
\newcommand{\no}{\nonumber}

\begin{document}
\title{Threshold theorem in isolated quantum dynamics with stochastic control errors}
\author{Manaka Okuyama$^1$}
\author{Kentaro Ohki$^2$}
\author{Masayuki Ohzeki$^{1,3,4}$}
\affiliation{$^1$Graduate School of Information Sciences, Tohoku University, Sendai 980-8579, Japan}
\affiliation{$^2$Graduate School of Informatics, Kyoto University, Kyoto 606-8501, Japan}

\affiliation{$^3$Department of Physics, Tokyo Institute of Technology, Oh-okayama, Meguro-ku, Tokyo,152-8551, Japan}
\affiliation{$^4$Sigma-i Co., Ltd., Tokyo 108-0075, Japan} 

\begin{abstract} 
We investigate the effect of stochastic control errors in the time-dependent Hamiltonian on isolated quantum dynamics.
The control errors are formulated as time-dependent stochastic noise in the Schr\"odinger equation.
For a class of  stochastic control errors, we establish a threshold theorem that provides a sufficient condition to obtain the target state, which should be determined in noiseless isolated quantum dynamics, as a relation between the number of measurements  and noise strength.
The theorem guarantees that if the sum of the noise strengths is less than the inverse of computational time, the target state can be obtained through a constant-order number of measurements.
If the opposite is true, the number of measurements to guarantee obtaining the target state increases exponentially with computational time. 
Our threshold theorem can be applied to any isolated quantum dynamics such as quantum annealing and adiabatic quantum computation.


\end{abstract}
\date{\today}
\maketitle

\section{Introduction}
Recent advances in experimental techniques have enabled the experimental realization of quantum dynamics, and it has become increasingly important to understand quantum dynamics.
In particular, recent efforts to realize quantum computation are progressing rapidly~\cite{Johnson,Arute}; hence, the precise control of quantum dynamics is required.

Although quantum dynamics is ideally described by the Schr\"odinger equation, the influence of the external environment cannot be ignored in experimental systems.
The external environment has two main effects on quantum systems: the influence from a heat bath and control errors of the Hamiltonian.
Here, we consider a case in which the influence of the heat bath can be eliminated.
Therefore, the dynamics of the target quantum system can be realized if the Hamiltonian can be controlled in an ideal manner.
However, it is difficult to control the Hamiltonian without errors in experimental systems.
Then, a natural question arises: is isolated quantum dynamics robust to the effects of control errors?
If the properties of isolated quantum dynamics dramatically change because of control errors, then it will be difficult to control the target quantum system in experimental systems, even if we can eliminate the effect of the heat bath.
For example, quantum annealing~\cite{KN,RCC,BBRA,BRA,SMTC} or adiabatic quantum computation~\cite{FGGS,ADKLLR,MLM} utilizes quantum dynamics for computation, but the theory of quantum error correction and suppression~\cite{JFS,Lidar,QL,PAL,VMKOBS} that has been established is not as complete as that of the circuit model~\cite{Shor,NC,Preskill}.
Therefore, it is vital to investigate the influence of control errors on isolated quantum dynamics from the perspective of analog quantum computation.

There are two main types of control errors that can occur in the time-dependent Hamiltonian.
One is time-invariant noise \cite{MAL,MGA,RC}, which acts as a bias. 
This type of errors modifies the Hamiltonian and will not be discussed in the present study.
The other, which we will focus on, is stochastic control noise \cite{WZ,GZ,DRC,ACHQGHLG}.
We formulate stochastic noise in unitary dynamics as time-varying stochastic noise.
The time evolution of the system is described by the stochastic differential equation~\cite{Gardiner}.
The present study examines whether it is possible to obtain the target state, which should be determined in noiseless time evolution, in the presence of stochastic control errors.
For this purpose, we establish a threshold theorem that provides a sufficient condition for obtaining the target state in the Schr\"odinger equation with stochastic control errors.
Our threshold theorem clarifies that the number of measurements to guarantee obtaining the target state strongly depends on the computational time and noise strength.

\section{Threshold theorem in isolated quantum dynamics with stochastic control errors}
We consider the following isolated quantum dynamics:
\be
i \frac{d}{dt} |\psi(t)\rangle&=& \hat{H}(t) |\psi(t)\rangle, \label{noise-less}
\ee
where $0\le t\le T$ and $\hbar=1$.
By using the measurement basis $|n\rangle$, we expand the final state $|\psi(T)\rangle$ as
\be
|\psi(T)\rangle&=& \sum_n C_n|n\rangle.
\ee
In the following, the entire derivation is based on projective measurements.
We are interested in the $m$th eigenstate $|m\rangle$ of the measurement basis, and its probability amplitude $C_m$ at the final state $|\psi(T)\rangle$ is given by
\be
C_m = 1-\epsilon ,
\ee
where $0<\epsilon<1$ (without loss of generality, we have adjusted the global phase of $|\psi(T)\rangle$ so that $C_m $ is a positive real number).
Then, if the number of measurements $r$ satisfies 
\be
r \gg \frac{1}{(1-\epsilon)^2},
\ee
we succeed in obtaining the target state $|m\rangle$.

However, it is difficult to completely control the time-dependent Hamiltonian $\hat{H}(t)$ without encountering control errors in experimental systems.
We incorporate the control errors of $\hat{H}(t)$ into the Schr\"odinger equation as noise that occurs stochastically at each moment.
Because we consider isolated quantum dynamics, control errors should also be described as a unitary time evolution.
It is well known that norm-preserving stochastic noise can be described by the Stratonovich process~\cite{WZ,GZ}.
Thus, we express the Schr\"odinger equation with stochastic noise as follows:
\be 
i d |\phi(t)\rangle&=& \left( \hat{H}(t)dt   + \sum_{k}\hat{H}_{{\rm error},k}(t)  \circ dW_k(t)  \right) |\phi(t)\rangle, \label{sto-Sch}
\ee
where $k$ is an index of stochastic control errors, $\hat{H}_{\rm error,k}(t)$ describes a stochastic control error, $W_k(t)$ describes standard Brownian motion, and the symbol ``\ $\circ$" denotes the Stratonovich interpretation. 
The equivalent Ito process is given by
\be 
i d |\phi(t)\rangle&=& \left( \hat{H}(t)dt  + \sum_{k}\hat{H}_{{\rm error},k}(t)  \bullet dW_k(t)  \right) |\phi(t)\rangle
\no\\
&&- \frac{i}{2} \sum_{k}\hat{H}_{{\rm error},k}^2(t)  |\phi(t)\rangle dt, \label{noisy-qa}
\ee
where the symbol ``\ $\bullet$" denotes the Ito interpretation. 
We emphasize that the norm of $|\phi(t)\rangle$ is always preserved in time evolution.

Furthermore, we assume that $\hat{H}_{{\rm error},k}(t)$ satisfies the following error condition
\be
\hat{H}_{{\rm error},k}^2(t) &=& g_k^2(t) \hat{I}, \label{local-cond}
\ee
where $\hat{I}$ is the identity operator and $g_k(t)$ is an arbitrary time-dependent function. 
We note that a system composed of qubits naturally satisfies Eq. (\ref{local-cond}). 
For example, when $\hat{H}_{{\rm error},k}(t)$ is constructed from the product of the Pauli operators at several neighboring, $\hat{H}_{{\rm error},k}(t)=g(t)\sigma_1^x \sigma_2^x\cdots \sigma_i^x$, the error condition (\ref{local-cond}) is satisfied.
On the other hand, this condition is not satisfied in a bosonic system.

Under these settings, we examine whether it is possible to obtain the target state.
If stochastic control errors have a devastating effect on isolated quantum dynamics, the target state cannot be obtained in experimental systems.
Additionally, if the number of measurements to guarantee obtaining the target state depends on the problem size $N$, stochastic control errors have a serious influence on the difficulty of the problem.
Our threshold theorem, which clearly addresses these issues, is described as follows.
\begin{theorem} \label{th2}
We are interested in the $m$th eigenstate $|m\rangle$ of the measurement basis and its probability amplitude $C_m$ at the final time $T$ in the noiseless Schr\"odinger equation (\ref{noise-less}) is given by
\be
C_m=1-\epsilon,
\ee
where $0<\epsilon<1$.
If $\epsilon$ and the noise strength $g_k(t)$ fulfill
\be
&&\epsilon + \delta +\alpha e^{\frac{1}{2}\int_0^T \sum_k g_k^2(t) dt}  \le 1, \label{main-ineq}
\ee
for $0<\delta<1-\epsilon$ and $\alpha>0$, then the $r$ measurement outcomes in the Schr\"odinger equation with stochastic control errors (\ref{noisy-qa}) satisfy
\be
\frac{1}{r}\sum_{i=1}^r |C_{i,m}|^2 > \alpha^2 , \label{sum-probability}
\ee
with probability greater than $1-\exp ( -r\delta^2/2 )$, where $C_{i,m}$ is the probability amplitude of the $m$th eigenstate at the final time $T$ in one realized time trajectory of Eq. (\ref{noisy-qa}).
\end{theorem} 
\begin{remark}
For $\exp ( -r\delta^2/2 )\ll 1$, Eq. (\ref{sum-probability}) holds with almost probability 1, then, if $r \alpha^2 \gg 1$ is satisfied, the target state $|m\rangle$ is always one of the $r$ measurement outcomes.
Thus, the casual form of the conditions to guarantee obtaining the target state is given by
\be
&&\epsilon + \delta +\alpha e^{\frac{1}{2}\int_0^T \sum_k g_k^2(t) dt}  \le 1, 
\\
&&\exp ( -r\delta^2/2 )\ll 1 ,
\\
&&r \alpha^2 \gg 1 .
\ee
\end{remark}
Our threshold theorem states that the number of measurements to guarantee obtaining the target state depends strongly on the strength of the noise.
For simplicity, we consider a case in which the strength of the noise is time-independent: $g_k(t)=g_k$.
Then, from Eq. (\ref{main-ineq}), the following condition must be satisfied for the number of measurements to be bounded by a constant order: 
\be
\frac{T}{2} \sum_k g_k^2 = \mathcal{O}(1) .
\ee
Thus, for any stochastic control errors satisfying Eq. (\ref{local-cond}), if the sum of the noise strengths is less than the inverse of the computational time, the target state can be obtained through a constant-order number of measurements.
Conversely, if the sum of the noise strengths is greater than the inverse of the computational time, the number of measurements to guarantee obtaining the target state increases exponentially with respect to the computational time.
In conclusion, stochastic control errors can have a serious impact on the difficulty of the problem, depending on the noise strength.

For example, we apply our threshold theorem to quantum annealing.
We consider a case in which the computational time $T$ is given by a polynomial of the problem size $N$: 
\be
T = \mathcal{O}(N^a),
\ee
which is efficiently solved by quantum annealing.
Then, we must suppress the noise as the problem size increases:
\be
\frac{1}{2}\sum_k g_k^2 =\mathcal{O}\left(N^{-a} \right).
\ee
Otherwise, from Eq. (\ref{main-ineq}), the number of measurements to guarantee obtaining the target state increases exponentially with respect to the problem size:
\be
r \gg e^{N^a \sum_k g_k^2} .
\ee
Therefore, when noise suppression fails, stochastic control errors may change an efficient quantum-annealing-based solution to the problem into an inefficient solution.

Before providing the proof, we emphasize that our threshold theorem is only a sufficient condition for any isolated quantum dynamics.
Incorporating the structure of the problem might improve our result.

\section{Proof of threshold theorem}
From Eqs. (\ref{noisy-qa}) and (\ref{local-cond}), the time evolution of the expectation of the state is described as 
\be
i \frac{d}{dt} \mathbb{E}\left[  |\phi(t)\rangle \right]&=& \left( \hat{H}(t) - \frac{i}{2} \sum_{k} g_k^2(t) \right) \mathbb{E}\left[  |\phi(t)\rangle \right] ,
\ee
where we used $\mathbb{E}[dW_k(t)  ]=0$ in the Ito process.
Then, using the fact that $|\psi(T)\rangle$ is the solution of Eq. (\ref{noise-less}), we find
\be
e^{\frac{1}{2}\int_0^T \sum_k g_k^2(t) dt}\mathbb{E}[ |\phi(T)\rangle]=|\psi(T)\rangle . \label{simple-relation}
\ee
We describe the state in one realized time trajectory of Eq. (\ref{noisy-qa}) as $|\phi_i(t)\rangle$ and expand it at the final time $T$ as
\be
|\phi_i(T)\rangle &=& \sum_n C_{i,n}|n\rangle .
\ee
From Eq. (\ref{simple-relation}), we immediately find 
\be
e^{\frac{1}{2}\int_0^T \sum_k g_k^2(t) dt} \mathbb{E}[C_{i,m}] &=&C_m=1-\epsilon.
\ee
Then, using the Chernoff--Hoeffding inequality~\cite{Chernoff,Hoeffding}, we have
\be
\textbf{Pr}\left[   e^{\frac{1}{2}\int_0^T \sum_k g_k^2(t) dt} \frac{1}{r}\sum_{i=1}^r \Re C_{i,m} -  (1-\epsilon) \le -\delta \right] \le \exp\left( \frac{-r\delta^2}{2} \right) ,   \label{Chernoff-Hoeffding}
\ee
where $C_{i,m}=\Re C_{i,m}+i  \Im C_{i,m}$, $\delta>0$, and we used $-1 \le \Re C_{i,m} \le1$.
In the following, we set $0<\delta<1-\epsilon$ and always consider the case in which
\be
  e^{\frac{1}{2}\int_0^T \sum_k g_k^2(t) dt} \frac{1}{r}\sum_{i=1}^r \Re C_{i,m}   >   1-\epsilon- \delta  . \label{necessary}
\ee
We note that this inequality plays an important role as a constraint and holds with probability greater than $1-\exp ( -r\delta^2/2 )$ from Eq. (\ref{Chernoff-Hoeffding}).

Next, we consider the case in which the probability amplitudes of the target state in the $r$ realized time trajectories of Eq. (\ref{noisy-qa}) satisfy 
\be
\frac{1}{r}\sum_{i=1}^r |C_{i,m}|^2 > \alpha^2 . \label{gene-success}
\ee
%
In the following, we prove that, under the conditions (\ref{main-ineq}) and (\ref{necessary}), Eq. (\ref{gene-success}) holds.
In order to accomplish this, we consider  
\be
\frac{1}{r}\sum_{i=1}^r |C_{i,m}|^2 \le \alpha^2 , \label{gene-fault}
\\
\alpha e^{\frac{1}{2}\int_0^T \sum_k g_k^2(t) dt} >1-\epsilon-\delta \label{main-ineq-inverse} ,
\ee
then, it is sufficient to prove that, under the condition (\ref{necessary}), Eq. (\ref{main-ineq-inverse}) is necessary for Eq. (\ref{gene-fault}) to hold.
Under Eq. (\ref{gene-fault}), $\sum_{i=1}^r  \Re C_{i,m} $ takes the maximum value $r\alpha$ when $\Re C_{i,m}=\alpha$ and $\Im C_{i,m}=0$.
Thus, for  Eq. (\ref{gene-fault}) to hold under the condition (\ref{necessary}), the following inequality must be satisfied:
\be
\alpha e^{\frac{1}{2}\int_0^T \sum_k g_k^2(t) dt} > 1-\epsilon-\delta, \no
\ee
which is just Eq. (\ref{main-ineq-inverse}).

In summary, Eq. (\ref{necessary}) holds with probability greater than $1-\exp ( -r\delta^2/2 )$ and Eq. (\ref{gene-success}) holds under the conditions (\ref{main-ineq}) and (\ref{necessary}), which is proof of the theorem.

\section{Conclusions}
We have established a threshold theorem that provides a sufficient condition for obtaining the target state in isolated quantum dynamics with any stochastic control errors satisfying Eq. (\ref{local-cond}).
Our threshold theorem guarantees that if the sum of the noise strengths is less than the inverse of the computational time, the target state can be obtained through a constant-order number of measurements.
However, if the sum of the noise strengths is larger than the inverse of the computational time, the number of measurements to guarantee obtaining the target state increases exponentially with respect to the computational time.

Furthermore, we imposed the error condition (\ref{local-cond}) on stochastic control errors.
If this condition is broken, the simple relation (\ref{simple-relation}) does not hold.
Then, we cannot guarantee that the target state can be obtained by increasing the number of measurements.
In other words, stochastic control errors that do not satisfy Eq. (\ref{local-cond}) have a serious influence  on isolated quantum dynamics.

In the present study, we considered only time-varying noise as a control error.
However, time-invariant noise can also be considered as a control error \cite{MAL,MGA,RC}, and our threshold theorem cannot be applied to such noise.
Time-invariant noise modifies the Hamiltonian.
In order to obtain the target state in experimental systems, such noise must be reduced to the limit, or a counterpart of our threshold theorem must be derived for such noise.

Finally, it is worth mentioning that we have considered only the sufficient condition to obtain the target state, and model-dependent properties may reduce the number of measurements required.
For example, in adiabatic quantum computation, adiabatic time evolution suppresses the diabatic transition from the ground state to other excited states.
In such cases, the effect of stochastic control errors may also be reduced.

The present work was financially supported by JSPS KAKENHI Grant No. 18H03303, 19H01095, 19K23418,  20H02168 and 21K13848 and by JST-CREST (No. JPMJCR1402) from the Japan Science and Technology Agency.



\end{document}